\documentclass[a4paper,11pt]{article}
\usepackage{pos}
\usepackage{booktabs}
\usepackage{amsmath}

\title{Searching for the QCD critical point with relativistic viscous hydrodynamics}

\author[a]{Kevin P. Pala}
\author[b]{Surkhab K. Virk}
\author[b]{Isabella Danhoni}
\author[c]{Fernando Gardim}
\author[d]{Christopher Plumberg}
\author*[b]{Jacquelyn Noronha-Hostler}

\affiliation[a]{Instituto de Física, Universidade de São Paulo, Rua do Matão 1371, 05508-090 São Paulo-SP, Brazil}

\affiliation[b]{The Grainger College of Engineering, Illinois Center for Advanced Studies of the Universe,
Department of Physics, University of Illinois at Urbana-Champaign, Urbana, IL 61801, USA}

\affiliation[c]{Instituto de Ciência e Tecnologia, Universidade Federal de Alfenas, 37715-400 Poços de Caldas, MG,
Brazil}

\affiliation[d]{Natural Science Division, Pepperdine University, Malibu, CA 90263, USA}

\emailAdd{kevin.p.pala@gmail.com}

\abstract{With the STAR's Beam Energy Scan II, a rich amount of data exist to search for the QCD critical point. However, a central challenge is that the critical point cannot be directly extracted from data but rather is probed dynamically in an expanding, cooling system that may be far-from-equilibrium. In this talk, we show the latest results from equations of state with a movable critical point generated by the MUSES collaboration, run in the 3+1D relativistic viscous hydrodynamics code CCAKE. We discuss numerical improvements to CCAKE that include ``glass'' initial conditions and adaptive smoothed particle hydrodynamics, allowing us to rapidly and accurately obtain numerical solutions to the hydrodynamic equations of motion.  We can quantify the interplay between beam energy, location of the critical point, and rapidity range. }

\FullConference{43rd International Conference on High Energy Physics (ICHEP 2026)\\
30 July  to 5 August , 2026\\
Natal, Brazil\\}

\begin{document}
\maketitle


\section{Introduction}

While we know that the phase transition from deconfined quarks and gluons (known as the quark gluon plasma) into  hadrons is a cross-over at vanishing net-baryon densities \cite{Aoki:2006we}, it has long been suspected that a critical point can occur  at large chemical potentials $\mu_B$ followed by a first-order line \cite{Stephanov:1998dy}.
The STAR collaboration at RHIC studied a wide range of center of mass beam energies $\sqrt{s_{NN}}$  known as the Beam Energy Scan  II (BESII). 
Since lower $\sqrt{s_{NN}}$ leads to less Lorentz contraction of nuclei, the nuclei take longer to pass through each other during the very initial stages of the collision. 
The longer time allows for baryons to be stopped within the collision such that it ``dopes'' the quark gluon plasma with a larger net-baryon density, $n_B$. 
By varying $\sqrt{s_{NN}}$, we can systematically vary the $n_B$.  
At a critical point, we expect that the correlation length $\xi$ to grow enormously. Thus, a key potential signature of the QCD critical point is net-proton fluctuations \cite{STAR:2020tga,STAR:2025zdq}, which  BESII has measured with more  fixed target data anticipated. There was also an interesting dip in the mean transverse momentum $\langle p_T\rangle$ fluctuations reported  \cite{2xsn-rgx3} from the STAR fixed program. 
These data  requite theoretical input to understand the influence of the QCD critical point. 

Theoretical frameworks have advanced enormously in the past decade to make more direct model-to-data comparisons at these $\sqrt{s_{NN}}$. Equations of state (EOS) from lattice QCD expansions can now reach out to about $\mu_B/T\sim 3.5$ \cite{Borsanyi:2021sxv}, and alternative methods have been developed that reproduce lattice QCD and can extend beyond it. Relativistic viscous fluid dynamics now contain the full BSQ (baryon number, strangeness, and electric charge) diffusion matrix \cite{Pala:2025qoa}, implemented within open-source software \cite{ccakesite}, although further work is still required on critical fluctuations in these frameworks, (see ongoing work e.g. \cite{Bzdak:2019pkr,An:2021wof}). 
We discuss the available MUSES  EOS  \cite{Jahan:2026hvs} and demonstrate its compatibility with the dynamical framework of the NuclearConfectionery~\cite{Pala:2025qoa}.

\section{Simulations setup}

\begin{figure}[h]
    \centering
    \begin{tabular}{cc}
      \includegraphics[width=0.48\linewidth]{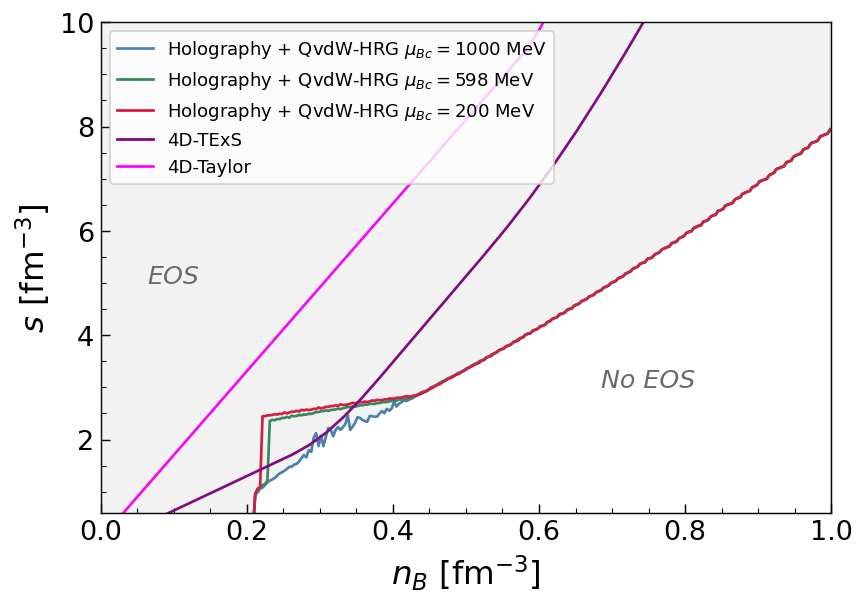}    &  \includegraphics[width=0.48\linewidth]{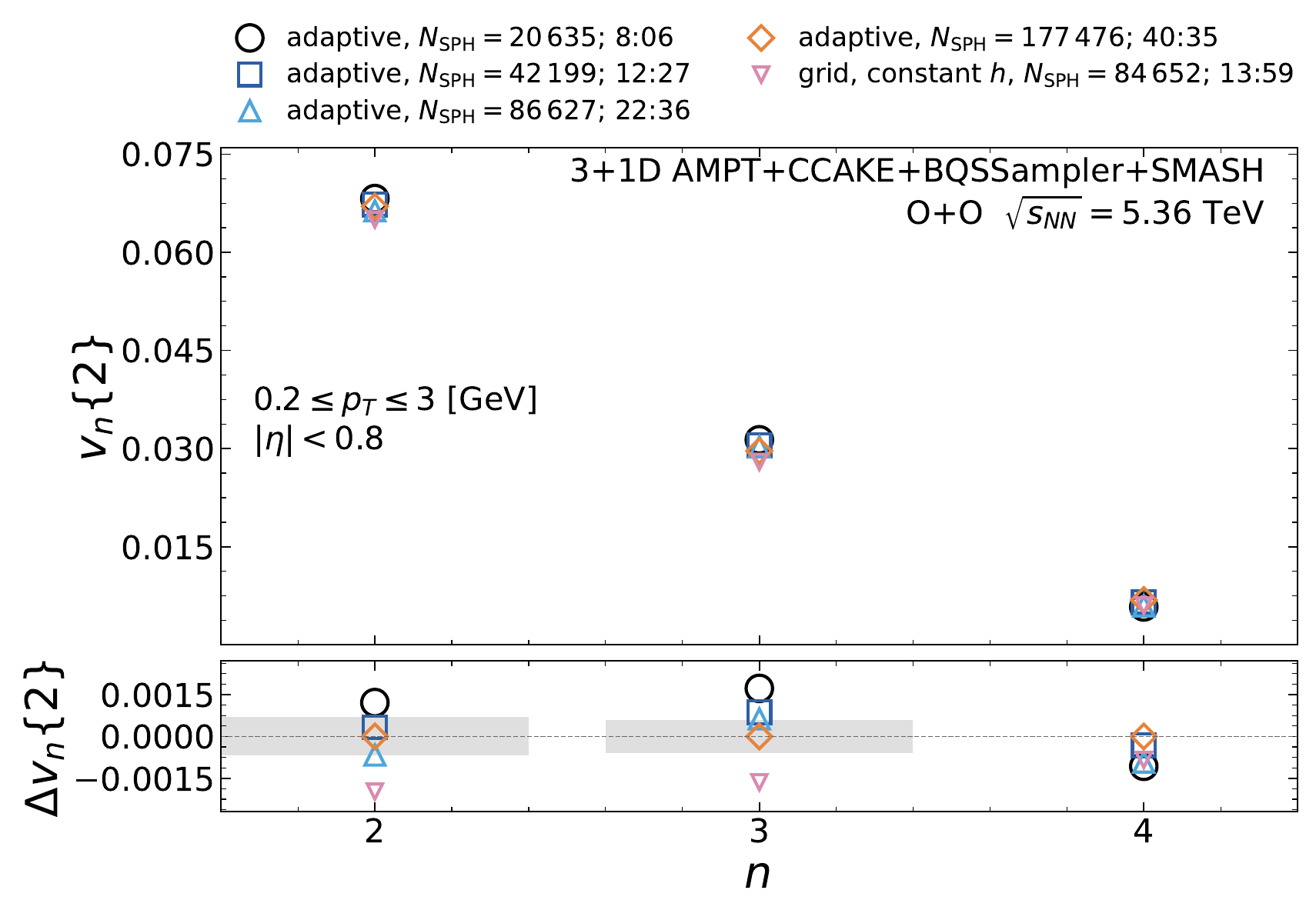}
    \end{tabular}
   
    \caption{Left: Range of the various MUSES EOS in the natural hydrodynamic variables of CCAKE. Figure from \cite{Jahan:2026hvs}. Right: Convergence test for ``glass'' initial conditions and adaptive SPH for O-O 5.36 TeV. The run time (minutes:seconds) is shown for the same initial condition, in each specific setup. 
    }
    \label{fig:snb}
\end{figure}

\textit{MUSES EOS:} The MUSES collaboration finished the \textit{Calliope} release \cite{Jahan:2026hvs} that developed open-source EOS modules relevant for heavy-ion collision.   \textit{Synthesis} merged different EOS modules across overlapping regimes of validity, allowing for cross-overs, critical points, and first-order phase transitions. 
The output from \textit{Synthesis} is in the grand canonical ensemble basis i.e. temperature and baryon chemical potential $T,\mu_B$ whereas NuclearConfectionery simulations require entropy and net-baryon density $s,n_B$. 
The \textit{EOS Inverter} module converts a 1D-4D EOS in $T,\mu_B$ into $s,n_B$. 
In Fig.\ \ref{fig:snb} left is the range of coverage of the available MUSES EOS. The gap in the bottom right demonstrates the limitation of existing EOS.  We fill in this gap with ``back-up'' EOS that are conformal, but ensure thermodynamic consistency \cite{Plumberg:2024leb}. 
We focus in this proceedings only on the merged Holography+QvdW-HRG
table with $\mu_S=\mu_Q=0$ (however, see \cite{Li:2026lxx}) because is has a movable critical point and  the widest coverage in $T,\mu_B$.  
We included the metastable and unstable regions of the EOS, to account for the jump in $n_B$ across the first-order phase transition, but in our simulations we never obtain a negative pressure.

\textit{Adaptive Smoothed Particle Hydrodynamics across $\sqrt{s_{NN}}$:} The NuclearConfectionery \cite{Pala:2025qoa} covers the initial state (AMPT \cite{Zhang:1999bd}), relativistic viscous hydrodynamics (CCAKE \cite{Plumberg:2024leb,Pala:2025qoa}), sampling of a fluid into hadrons (BQSSampler  \cite{Pala:2025qoa}), and hadronic interactions (SMASH \cite{SMASH:2016zqf}). 
The algorithm from CCAKE relies on smoothed particle hydrodynamics (SPH), which is a common numerical tool used in astrophysics \cite{Rosswog:2014dga} and was first used in heavy-ion collisions with the NEXSPheRIO code \cite{Aguiar:2000hw}. The fluid relies on a Lagrangian method that follows the path of individual SPH ``particles''. 

Two upgrades have improved both numerical accuracy and efficiency. 
The initial conditions are not initialized on a fixed grid across $\Delta x, \Delta y, \Delta z$, but instead sampled from the local energy density. Then, the initial condition each  SPH particles is given a fictitious repulsive force and allowed to settle into an optimal initial configuration.  
Such an approach is known as WVT-relaxation, which produces the WVT-glass\footnote{This glass has nothing to do with glasma initial conditions.} initial conditions \cite{Diehl:2012bt,Kubli:2025aqs}  because it leads to an amorphous, isotropic distribution with no preferred direction.
Studies have found that glass initial conditions do a significantly better job of resolving finer features of simulations \cite{Diehl:2012bt}. 
The other improvement is the incorporation of adaptive SPH ~\cite{Springel:2001qb,Price:2010hv}.  SPH has an underlying smoothing scale called $h$ that sets the minimum resolution scale of the fluid. 
In previous SPH codes for heavy-ion collisions, $h$ was held constant throughout simulations. 
However, by allowing adaptive $h$ where the resolution scale varies by the local SPH particle density, it sets a low resolution for the edges of the fluid that do not significantly affect experimental observables and a small $h$ at the center of the fluid that contains finer structures. As such, one requires significantly fewer SPH particles for high numerical accuracy and can better resolve rapid expansions. 
We find that incorporating both glass initial conditions and adaptive SPH improves our energy conservation in CCAKE by an order of magnitude. Because we need significantly fewer SPH particles, the run time is shorter (approximately 1 minute less for O-O collisions). In Fig.\ \ref{fig:snb} right, results used the parametrization in Tab.\ \ref{tab:oo_params}

\begin{table}[htbp]
  \centering
  \caption{Simulation parameters for O+O collisions.}
  \label{tab:oo_params}
  \begin{tabular}{l c @{\hspace{2.5em}} l c @{\hspace{2.5em}} l c}
    \toprule
    $K$                        & 0.749 & $\eta T/w$                     & 0.14 &  $\tau_0$ [fm/$c$]          & 0.4\\
    $\sigma_r$ [fm]            & 1.25  & $\zeta T/w$                    & $1.4\,(1/3 - c_s^2)^2$ & \\
    $\sigma_\eta$              & 1.40  & $e_{\mathrm{fo}}$ [GeV/fm$^3$] & 0.257     & \\
    \bottomrule
  \end{tabular}
\end{table}

\section{Results}

\begin{figure*}[!ht]
    \centering
    \includegraphics[width=0.6\linewidth]{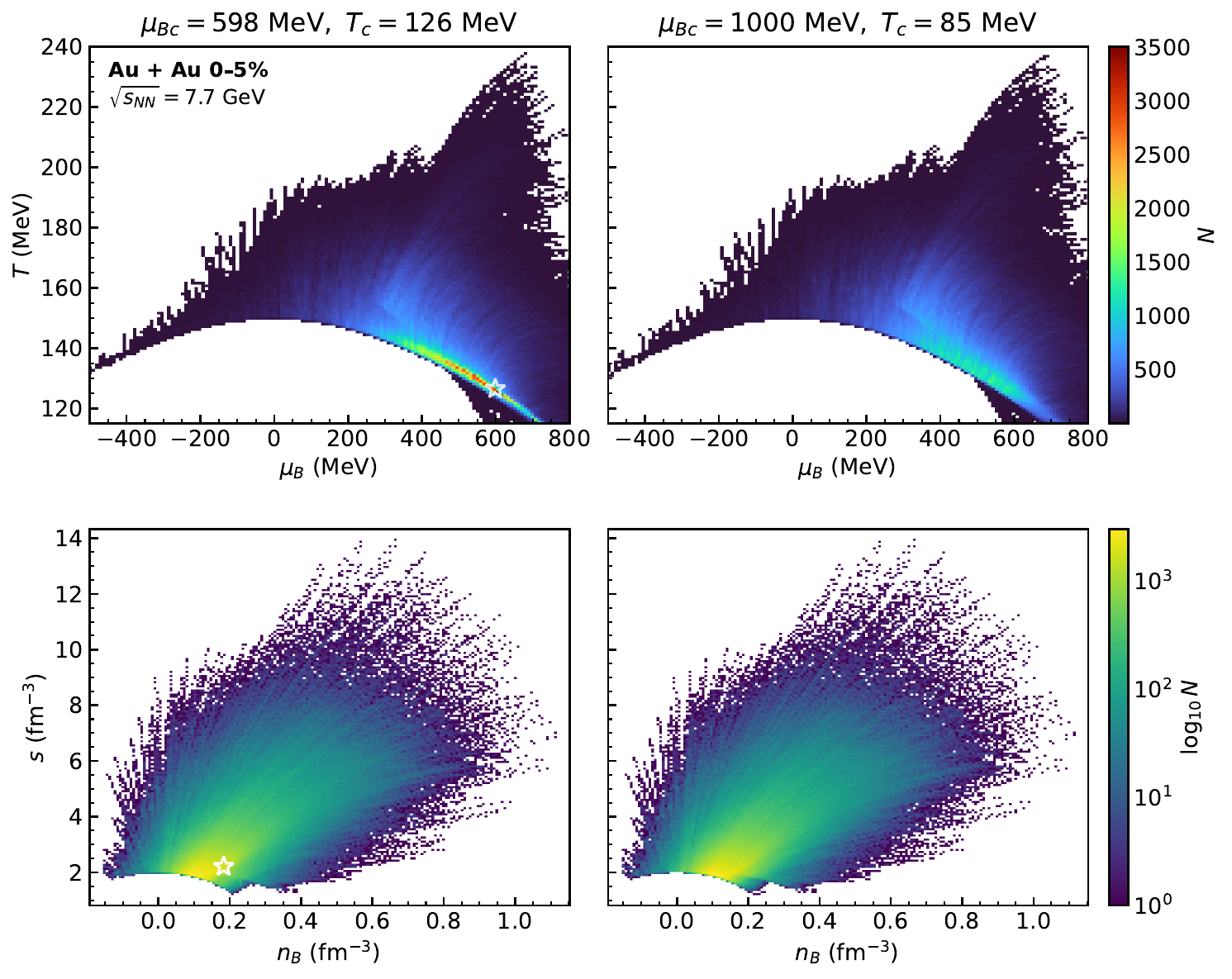}
    \caption{Passage through the QCD phase diagram for a single initial condition. We used the Holography+QvdW-HRG EoS with 2 different critical point positions.  Taken from \cite{Jahan:2026hvs}. 
    }
    \label{fig:holotrajectories}
\end{figure*}

Using our framework, we can now study the passage through the QCD phase diagram at  $\sqrt{s_{NN}}=7.7$ GeV, one of the beam energies of significant interest. 
We specifically choose $\sqrt{s_{NN}}=7.7$ GeV because it has a large enough reach in $\mu_B$ that a subset of the fluid passes through the critical point, if it is at $\mu_B^c=598$ MeV (the location predicted from the holography EOS \cite{Hippert:2023bel} when constrained by lattice QCD). 
In Fig.\ \ref{fig:holotrajectories}, we see the amount of SPH particles that appear at each point in the QCD phase diagram across all time steps until freeze-out is reached. 
The way to understand these figures is that regimes with the most hits (i.e. the red/orange/yellow regions) are where the system spends the most time.  

Here we compare one EOS with the critical point at $\mu_B^c=598$ MeV to another where the critical point is out of reach of these beam energies i.e. $\mu_B^c=1000$ MeV
Thus, from Fig.\ \ref{fig:holotrajectories}, we see that in the $T,\mu_B$ plane there is a significantly large amount of time spent in the fluid close to the critical point, when it is present.  In contrast, when the critical point is beyond the reach of that particular beam energy, significantly less time is spent at that location, which may be understood in terms of (dynamical) critical lensing \cite{Dore:2022qyz}.
Interestingly enough, the natural hydrodynamic variables are nearly insensitive to the location of the critical point. If we look at the $s,n_B$ plane in Fig.\ \ref{fig:holotrajectories}, we observe almost no difference between the distribution of the fluid cells.

\begin{figure}[h]
    \centering
    \includegraphics[width=\linewidth]{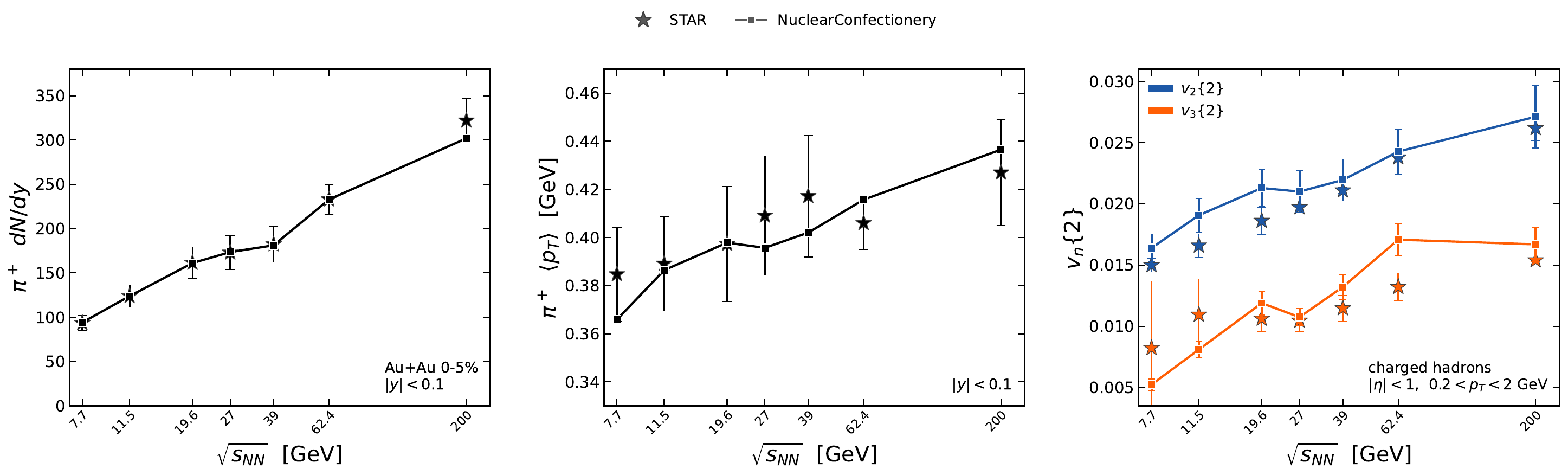}
    \caption{Comparisons of Au-Au collisions using NuclearConfectionery with STAR BESII data. }
    \label{fig:observables}
\end{figure}

In Fig.\ \ref{fig:observables} we show pion multiplicity across $\sqrt{s_{NN}}$ on the left, pion mean transverse momentum on center and the azimuthal anisotropies for charged particles $v_n\left\{2\right\}$ on the right.  Here we have small variations in the overall normalization by beam energy to reproduce these results.  Additionally, our shear viscosity currently has a $\mu_B$ dependence as well, such that it grows at larger $\mu_B$.  
We find that our results with a critical point are able to reproduce the STAR data reasonably well. The parametrization can be found in Table~\ref{tab:sim_params}.

\begin{table}[htbp]
  \centering
  \caption{Simulation parameters for Au+Au collisions at the BES energies:
    $K$ the overall normalization,
    $\sigma_r$ and $\sigma_\eta$ the transverse and longitudinal smearing widths,
    $\tau_0$ the state time,  $t_{\mathrm{cross}}$ the nuclear
    crossing time,  $w = e + P$ the enthalpy density, and the
    particlization criterion $e_{\mathrm{fo}} =0.2567$ [GeV/fm$^3$], common to all energies. Partons formed after $\tau_0$ are either fed as source terms or back-propagated onto the $\tau_0$ surface.
 }
  \label{tab:sim_params}
  \begin{tabular}{l ccccccc}
    \toprule
    $\sqrt{s_{NN}}$ [GeV] & 7.7 & 11.5 & 19.6 & 27 & 39 & 62.4 & 200 \\
    \midrule
    $K$                              & 0.8796 & 0.7685 & 1.0120 & 0.8007 & 1.0033 & 0.9722 & 1.000 \\
    $\sigma_r$ [fm]                  & 0.700  & 1.020  & 1.300  & 1.020  & 1.004  & 1.120  & 1.300 \\
    $\sigma_\eta$                    & 1.500  & 1.000  & 1.000  & 1.450  & 1.148  & 1.404  & 1.400 \\
    $\tau_0$ [fm/$c$]                & 1.70   & 1.50   & 1.50   & 1.47   & 1.19   & 0.81   & 0.50  \\
    Dynamical initialization               & on    & on    & on    & off   & off  & off  & off  \\
    \midrule
    $\eta T/w$ & \multicolumn{7}{c}{$\displaystyle
      \begin{cases}
        0.1014 + (0.169 - 0.1014)\,\dfrac{\mu_B}{0.2\ \mathrm{GeV}},
          & 0 \le \mu_B \le 0.2\ \mathrm{GeV} \\[8pt]
        0.169 + (0.390 - 0.169)\,\dfrac{\mu_B - 0.2\ \mathrm{GeV}}{0.2\ \mathrm{GeV}},
          & 0.2 < \mu_B < 0.4\ \mathrm{GeV} \\[8pt]
        0.390, & \mu_B \ge 0.4\ \mathrm{GeV}
      \end{cases}$} \\[24pt]
    $\zeta T/w$   & \multicolumn{7}{c}{$2.6\,(1/3 - c_s^2)^2$} \\
    \bottomrule
  \end{tabular}
\end{table}
\section{Conclusions and Outlook}

Here we discussed the upgrades and numerical developments within the NuclearConfectionery in order to make direct comparisons to experimental observables at RHIC's Beam Energy Scan.  Using the newly developed MUSES EOS, we were able to explore the reach of a single event at $\sqrt{s_{NN}}=7.7$ GeV both with and without a critical point to determine the amount of time spent close to the critical point.  Additionally, we find that our setup can provide a reasonable description of multiplicity and flow observables, making it suitable for future calculations of experimental observables and more model-to-data comparisons.  

\textit{Acknowledgments } This research was supported by the US-DOE Nuclear Science Grant No. DE-SC0023861,
within the framework of the Saturated Glue (SURGE) Topical Theory Collaboration, and
by the NSF within the framework of the MUSES collaboration, under grant number OAC-2103680, partially by from Fundação de Amparo à
Pesquisa do Estado de São Paulo (grants 2020/15893-4, 2024/08903-4 and 2018/24720-6), and supported by CNPq  through
307806/2025-1. 
This work used the Delta system
at NCSA
through allocation PHY250117 from the Advanced Cyberinfrastructure Coordination Ecosystem: Services and
Support (ACCESS) program, which is supported by NSF grants \#2138259, \#2138286,
\#2138307, \#2137603, and \#2138296.

\bibliographystyle{elsarticle-num}
\bibliography{inspire,NOTinspire}

@article{2xsn-rgx3,
  author = {al., B. E. Aboona et},
  journal = {Phys. Rev. Lett.},
  pages = {},
  year = {2026},
  month = {Jul},
  publisher = {American Physical Society},
}

@misc{ccakesite,
 year = {2024},
publisher = {GitHub},
    journal = {GitHub repository},
    note         = {\url{https://the-nuclear-confectionery.github.io/ccake-site/}}
}

@article{Aoki:2006we,
    author = "Aoki, Y. and others",
    journal = "Nature",
    volume = "443",
    pages = "675--678",
    year = "2006"
}

@article{Zhang:1999bd,
    author = "Zhang, Bin and others",
    journal = "Phys. Rev. C",
    volume = "61",
    pages = "067901",
    year = "2000"
}

@article{Kubli:2025aqs,
    author = "Kubli, Noah and Fothers",
    journal = "Astrophys. J. Lett.",
    volume = "999",
    number = "2",
    pages = "L40",
    year = "2026"
}

@article{Stephanov:1998dy,
    author = "Stephanov, Misha A. and Rajagopal, K. and Shuryak, Edward V.",
    journal = "Phys. Rev. Lett.",
    volume = "81",
    pages = "4816--4819",
    year = "1998"
}

@article{STAR:2020tga,
    author = "Adam, J. and others",
    collaboration = "STAR",
    journal = "Phys. Rev. Lett.",
    volume = "126",
    number = "9",
    pages = "092301",
    year = "2021",
    note = "[Erratum: Phys.Rev.Lett. 134, 139902 (2025)]"
}

@article{STAR:2025zdq,
    author = "Aboona, B. E. and others",
    collaboration = "STAR",
    journal = "Phys. Rev. Lett.",
    volume = "135",
    number = "14",
    pages = "142301",
    year = "2025"
}

@article{Borsanyi:2021sxv,
    author = "Bors{\'a}nyi, S. and others",
    journal = "Phys. Rev. Lett.",
    volume = "126",
    number = "23",
    pages = "232001",
    year = "2021"
}

@article{Pala:2025qoa,
    author = "Pala, Kevin P. and others",
    eprint = "2511.22852",
    archivePrefix = "arXiv",
    primaryClass = "nucl-th",
    month = "11",
    year = "2025"
}

@article{Bzdak:2019pkr,
    author = "Bzdak, Adam and others",
    journal = "Phys. Rept.",
    volume = "853",
    pages = "1--87",
    year = "2020"
}

@article{An:2021wof,
    author = "An, Xin and others",
    journal = "Nucl. Phys. A",
    volume = "1017",
    pages = "122343",
    year = "2022"
}

@article{Jahan:2026hvs,
    author = "Jahan, Johannes and others",
    eprint = "2606.26326",
    archivePrefix = "arXiv",
    primaryClass = "nucl-th",
    month = "6",
    year = "2026"
}

@article{Plumberg:2024leb,
    author = "Plumberg, Christopher and others",
journal = "Phys. Rev. C",
    volume = "111",
    number = "4",
    pages = "044905",
    year = "2025"
}

@article{Li:2026lxx,
    author = "Li, Zhibin and Li, Danning and Huang, Mei",
    eprint = "2607.19149",
    archivePrefix = "arXiv",
    primaryClass = "hep-ph",
    month = "7",
    year = "2026"
}

@article{SMASH:2016zqf,
    author = "Weil, J. and others",
    collaboration = "SMASH",
    journal = "Phys. Rev. C",
    volume = "94",
    number = "5",
    pages = "054905",
    year = "2016"
}

@article{Rosswog:2014dga,
    author = "Rosswog, Stephan",
    journal = "Liv. Rev. Comput. Astrophys.",
    volume = "1",
    number = "1",
    pages = "1",
    year = "2015"
}

@article{Aguiar:2000hw,
    author = "Aguiar, C. E. and others",
    journal = "J. Phys. G",
    volume = "27",
    pages = "75--94",
    year = "2001"
}

@article{Diehl:2012bt,
    author = "Diehl, Steven and others",
    journal = "Publ. Astron. Soc. Austral.",
    volume = "32",
    pages = "e048",
    year = "2015"
}

@article{Springel:2001qb,
    author = "Springel, Volker and Hernquist, Lars",
    journal = "Mon. Not. Roy. Astron. Soc.",
    volume = "333",
    pages = "649",
    year = "2002"
}

@article{Price:2010hv,
    author = "Price, Daniel J.",
    journal = "J. Comput. Phys.",
    volume = "231",
    pages = "759--794",
    year = "2012"
}

@article{Hippert:2023bel,
    author = "Hippert, Mauricio and others",
    journal = "Phys. Rev. D",
    volume = "110",
    number = "9",
    pages = "094006",
    year = "2024"
}

@article{Dore:2022qyz,
    author = "Dore, Travis and others",
    journal = "Phys. Rev. D",
    volume = "106",
    number = "9",
    pages = "094024",
    year = "2022"
}

\end{document}